\documentclass[review, 3p, twocolumn]{elsarticle}

\usepackage{graphicx}
\usepackage{bm}
\usepackage{hyperref}
\usepackage{siunitx}
\usepackage{miller}
\usepackage[version=4]{mhchem} 

\begin{document}

\begin{frontmatter}

\title{Image Difference Metrics for High-Resolution Electron Microscopy}

\author[1]{Manuel Ederer}
\ead{manuel.ederer@tuwien.ac.at}

\author[1]{Stefan L\"{o}ffler}
\address[1]{University  Service  Centre  for  Transmission  Electron  Microscopy, TU Wien, Wiedner Hauptstraße 8-10/E057-02, 1040 Wien, Austria}

\date{\today}

\begin{abstract}
\noindent
Digital image comparison and matching brings many advantages over the traditional subjective human comparison, including speed and reproducibility.
Despite the existence of an abundance of image difference metrics, most of them are not suited for high-resolution transmission electron microscopy (HRTEM) images.
In this work we adopt two image difference metrics not widely used for TEM images.
We compare them to subjective evaluation and to the mean squared error in regards to their behaviour regarding image noise pollution.
Finally, the methods are applied to and tested by the task of determining precipitate sizes of a model material. 
\end{abstract}

\begin{keyword}
image difference metric, transmission electron microscopy, high-resolution, SIFT, SSIM, MSE
\end{keyword}

\end{frontmatter}

\section{\label{Intro}Introduction}

With the ever growing amount of data electron microscopes and simulations are able to produce, automatic image evaluation becomes more important than ever.
Image evaluation can take many forms, from determining the best parameters or circumstances for observing certain structures in the image, to quantifying and measuring those structures. 
While the trend is moving away from manually analyzing images one by one, in many areas of electron microscopy it is still the norm.
Automatically evaluating high-resolution transmission electron microscopy (HRTEM) images, however, is drastically less time consuming
and brings the additional advantages of the result being reproducible and independent of the subjective user.
To this regard, there exist a number of image analysis methods such as for the structure analysis of carbonaceous materials \cite{palotas1996,sharma1999,Yehliu2011},
for the detection of displacements and lattice distances from HRTEM lattice images \cite{rosenauer1996, Bierwolf1993, Hytch1998},
for model-based quantification of high angle annular dark field images \cite{VanAert2012, DeBacker2016}
and various methods based on machine learning and neural networks \cite{Li2017,madsen2018,Ede2021}.
While these methods can yield excellent results, they are often only applicable to a very specific system or require large training sets.
A more general approach for characterizing an image is by its difference to a perfect reference image, calculated by a difference metric.
Unfortunately however, a universally applicable image difference metric also does not exist, only a multitude of methods with various advantages and disadvantages \cite{pedersen2009}. 
In HRTEM several image comparison methods are already in use, mainly for iterative digital image matching.
These methods include various image agreement factors, calculated with the cross-correlation factor \cite{Moebus1994},
the fractional mean absolute difference and discrimination factor \cite{smith1982}, or the $\chi^2$ goodness-of-fit test \cite{king1993}.
Usually, when simulated images are compared to the experiment using current methods, background noise is removed by applying Bragg filters to the Fourier-transformed images \cite{jong1989}.
The noise removal, however, in turn introduces new artifacts \cite{kauffmann2005}.
Thus, one of our main criteria to judge the usefulness of an image difference metric for HRTEM images 
is its robustness against noise.
However, robustness alone is useless when the difference metric in turn ignores fundamental changes in the image.
The, in theory, perfect difference metric should ignore changes caused by noise while still retaining its 
sensitivity to small changes of the underlying structure of the image.
We acknowledge that this criterion is to a great extent subjective as the concept of a defining structure of an image 
is ambiguous and largely dependent on the specific case which is the reason for the multitude of existing 
image difference metrics.
For HRTEM images, however, the image structure often coincides with the atomic structure of the material.
Thus, among other reasons, we have chosen our image metrics based on their ability to recognize position and orientation of
periodic shapes.\\
The first part of this work will deal with evaluating three promising image difference metrics 
by their response and robustness to different types of noise in the image.
We put a strong focus on shot noise as it is ever present in experimentally acquired images and even artificially 
added to simulated images when the goal is direct comparison,
while other sources of noise can be reduced by cooling the detector or use of direct detection.
In the second part of this work we apply the investigated difference metrics 
to the model task of automatically detecting precipitate sizes in simulated high-resolution images of \ce{Nb3Sn}.
While mainly serving as a further demonstration of the versatility of the chosen difference metrics,
the method can easily be extended to experimental applications where the automatic quantification of 
precipitates or similar structures is of interest.

\section{\label{Methods}Methods}

\begin{figure}
    \centering
    \includegraphics{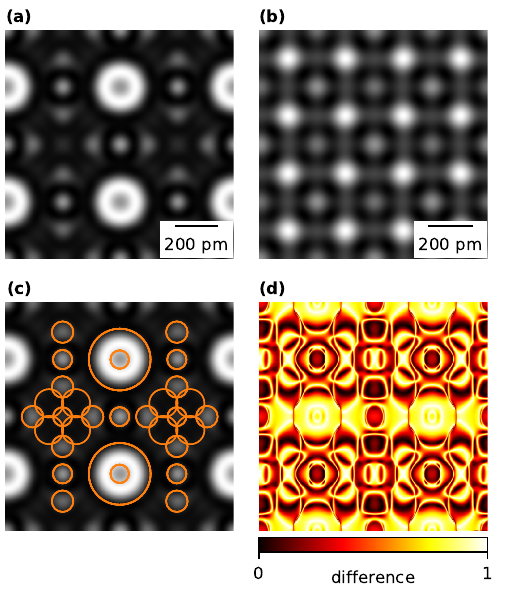}
    \caption{(a), (b) HRTEM simulations of \ce{Nb3Sn} and \ce{ZrO2} with a projected thickness of \SI{10.55}{\nano\meter} and \SI{10.18}{\nano\meter}, respectively.
      (c) The reference image from (a) with features detected and drawn using SIFT.
      Feature positions and sizes are indicated by orange circles.
      (d) Difference map between (a) and (b) calculated with cw-SSIM.
}
    \label{fig1}
\end{figure}

We use three different methods to gauge the difference of an image compared to a reference image:
the structural similarity index measure (SSIM) \cite{wang2004}, the scale invariant feature transform (SIFT) algorithm \cite{lowe1999} and the mean squared error (MSE).
SSIM and SIFT have been chosen based on their invariance to small translations, rotations, and noise 
while still showing sensitivity for small, local changes in the periodic structure.
MSE has been chosen because of its simplicity and prevalence as an image metric and serves as a standard.\\
The structural similarity index measure between a point $\mathbf{x}$ in image A and a
point $\mathbf{y}$ in image B is defined in the spatial domain as

\begin{equation}
  D(\mathbf{x}, \mathbf{y}) = 0.5 - \frac{(2\mu_x \mu_y + C_1)(2\sigma_{xy} + C_2)}{2(\mu_x^2 + \mu_y^2 + C_1)(\sigma_x + \sigma_y + C_2)},
\label{eq:SSIM}
\end{equation}
\noindent
where $\mu_x$ is the mean and $\sigma_x$ the variance of a circular image batch around $\mathbf{x}$, 
$\sigma_{xy}$ the covariance between the image batches around $\mathbf{x}$ and $\mathbf{y}$, and $C_1$ and $C_2$ are small, positive constants.
Note the difference in normalization compared to \cite{wang2004} to bring the index more in line
with the other employed methods and to indicate difference rather than similarity.
The total image difference between image A and image B is given by the average over all points.
In order to achieve more robustness against small translations, rotations, illumination changes and noise \cite{bayer2019, chang2000}, the similarity index is extended to the complex wavelet (cw) transform domain \cite{wang2005}
\begin{equation}
    \Tilde{D}(\mathbf{c_x}, \mathbf{c_y}) = 1 - \frac{2| \sum_{i=1}^N c_{x,i}c^*_{y,i} | + K}{\sum_{i=1}^N|c_{x,i}|^2 + \sum_{i=1}^N |c_{y,i}|^2 + K},
\label{eq:cwSSIM}
\end{equation}
\noindent
where $\mathbf{c_x} = \{ c_{x,i} | i = 1, ..., N \}$ and $\mathbf{c_y} = \{ c_{y,i} | i = 1, ..., N \}$ are sets of coefficients of the wavelet transform extracted at the same spatial location in the same wavelet subbands from images A and B, respectively, and K is a small, real, positive constant.
Fig.~\ref{fig1}(d) shows the complex wavelet SSIM (cw-SSIM) difference map between two unrelated example images. \\
In contrast to this method, the adapted scale invariant feature transform (SIFT) algorithm \cite{lowe1999} is based on automatic feature extraction of the reference image and the subsequent comparison of feature descriptors with the test image.
The features are detected by finding maxima and minima of a difference of Gaussian functions applied in scale space.
For this, the image is first smoothed by successive convolution with a Gaussian kernel and then resampled.
The differences between neighbouring levels in scale space are calculated and a pyramid in scale space is constructed.
Extremal points in this pyramid mark regions of high variation and are sufficient to characterize the image.
In Fig.~\ref{fig1}(c) orange circles mark the locations and sizes of key points in real space.
At each point for each level of the pyramid the image gradient magnitude and orientation are calculated.
The orientations are accumulated around each key point allowing us to assign it a canonical orientation.
For the key point descriptor a set of orientation histograms is created for a number of sub-regions around the key point, using the previously calculated gradients.
The orientation histograms are taken relative to the key point orientation ensuring invariance to global rotations and changes in illumination, respectively.
Key point descriptors at the same points in scale space of the test image are calculated in the same way
and subsequently the differences to the descriptors of the reference image are averaged resulting in a total difference between reference and test image.
Note that this method is mathematically speaking not a metric as the symmetry between the elements is lost in the last step, i.e. the difference depends on which image is the reference and which the test image.  
Nonetheless, the method is applicable and useful for all purposes where a distinct, optimal reference image is available.
A more in-depth description of the SIFT and the (cw-)SSIM methods is given in the supplementary information together with all parameters and constants used in the calculations.
\noindent \\
We define the normalized mean squared error (MSE) as
\begin{equation}
    \mathrm{MSE}(\mathbf{a}, \mathbf{b}) = \frac{1}{2} \sum_{i=1}^{N} \left( \frac{a_i}{\sqrt{\sum_{i=1}^N a_{i}^2}} - \frac{b_i}{\sqrt{\sum_{i=1}^N b_i^2}} \right) ^2
\end{equation}
\noindent
where $\mathbf{a} = \{ a_{i} | i = 1, ..., N; ~ a_{i} \geq 0\}$ and $\mathbf{b} = \{ b_{i} | i = 1, ..., N; ~ b_{i} \geq 0 \}$ are the sets of pixels 
of image A and B, respectively. 
This definition differs from the classic MSE in a few points.
First of all, the images are individually normalized before comparison using the Frobenius norm \cite{Golub1985}.
This is akin to normalizing TEM images by their electron dose.
Secondly, in order to make the MSE directly comparable to the other methods we normalize by the maximal achievable value, the sum of the individual norms squared, hence the factor $1/2$.\\
All HRTEM images in this work have been simulated based on the multi-slice algorithm \cite{Kirkland1998, Cowley1957}.
Unless stated otherwise, we have used an acceleration voltage of 80 kV and a collection semi-angle of \SI{30}{\milli\radian}.
For both of the investigated materials, \ce{Nb3Sn} and \ce{ZrO2}, only the [0 0 1] crystallographic direction is considered.

\subsection{Application to test cases}

\begin{figure*}
    \centering
    \includegraphics{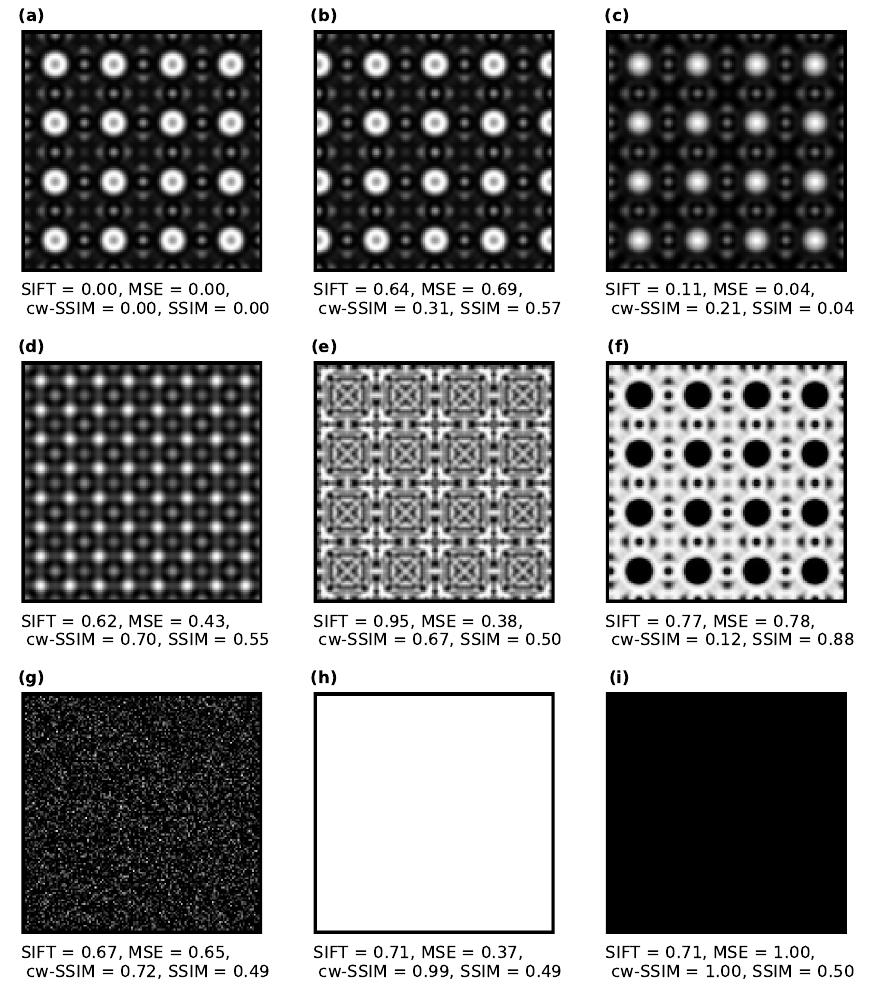}
    \caption{
    (a) Reference image: \ce{Nb3Sn} with a projected sample thickness of 21.1 nm. 
    (b) Horizontal shift of half a unit cell.
    (c) Defocus of -2 nm. 
    (d) \ce{ZrO2} with a projected sample thickness of 20.36 nm. 
    (e),(f) Images maximizing the SIFT or SSIM difference, respectively. 
    (f) Noise with a Gaussian distribution. (e) White image. (f) Black image.  
}
    \label{fig_bench}
\end{figure*}

A difference metric, especially when it is used to determine optimal theoretical and experimental parameter ranges, is only useful when the edge cases are known,
i.e. when one has a clear understanding how an image with 1, 0.5 or 0.25 difference to the reference image looks like.
A representative selection of images is shown in Fig.~\ref{fig_bench} together with the respective values calculated with the image difference metrics introduced earlier.
An HRTEM image of \ce{Nb3Sn} serves as reference image [Fig.~\ref{fig_bench}(a)] and demonstrates the only case where a difference metric can result in 0, i.e. when the test and reference image are the same.
A large translation of the image is shown in Fig.~\ref{fig_bench}(b).
All image difference metrics except for cw-SSIM result in a significant difference above 0.5 for this case, highlighting that proper alignment of the images is necessary when images with the same structure are compared. 
For Fig.~\ref{fig_bench}(c) a small defocus was applied to the original image.
This case shows the sensitivity of the individual methods to small but noticeable changes in the image.
Based on the general behaviour of the image difference metrics we choose a subjective threshold of 0.18 for SIFT, 0.16 for cw-SSIM and 0.07 for the MSE.
Below this threshold the difference is basically negligible and, thus,
image difference metrics can be used to find parameter ranges such as defocus where the resulting image is still acceptable.
Of note here is the strong indifference of the real space SSIM and the MSE and the strong sensitivity of the cw-SSIM relative to this type of image modulation.
Fig.~\ref{fig_bench}(d) with an HRTEM image of \ce{ZrO2} represents a case where both the images and the crystal lattices have an arbitrary different structure.
The resulting image differences highlight the fact that, while all the metrics have a theoretical maximum difference of 1, 
the difference for unconnected HRTEM images lies typically between 0.5 and 0.75 --- with the notable exception of the MSE.
The next two images show the results of Monte Carlo simulations with the goal of maximizing the image difference of a specific metric for the given reference image.
Fig.~\ref{fig_bench}(e) results in the largest possible distances of local ensembles of image gradient vectors in each feature of the respective images, thus maximizing the metric based on SIFT.
Fig.~\ref{fig_bench}(f) closely resembles the contrast reversal of the reference image with a few exceptions and results in the highest real space SSIM image difference possible for the given reference image. 
While the MSE and the SIFT method also result in a high image difference, the cw-SSIM method results in negligible difference, lower even than the defocus case. 
This highlights the fact that, while in most cases cw-SSIM is preferable to its real space counterpart, for defocus series and other image sets where a contrast reversal is possible, any of the other methods is a better choice than cw-SSIM. 
The third row of test images in Fig.~\ref{fig_bench} depicts how the difference metrics react to Gaussian noise and to a completely white or black image, the latter resulting in maximum image difference for the MSE and cw-SSIM methods.

\section{\label{ResDisc}Results and discussion}

\subsection{Influence of Noise}

\begin{figure*}
    \centering
    \includegraphics{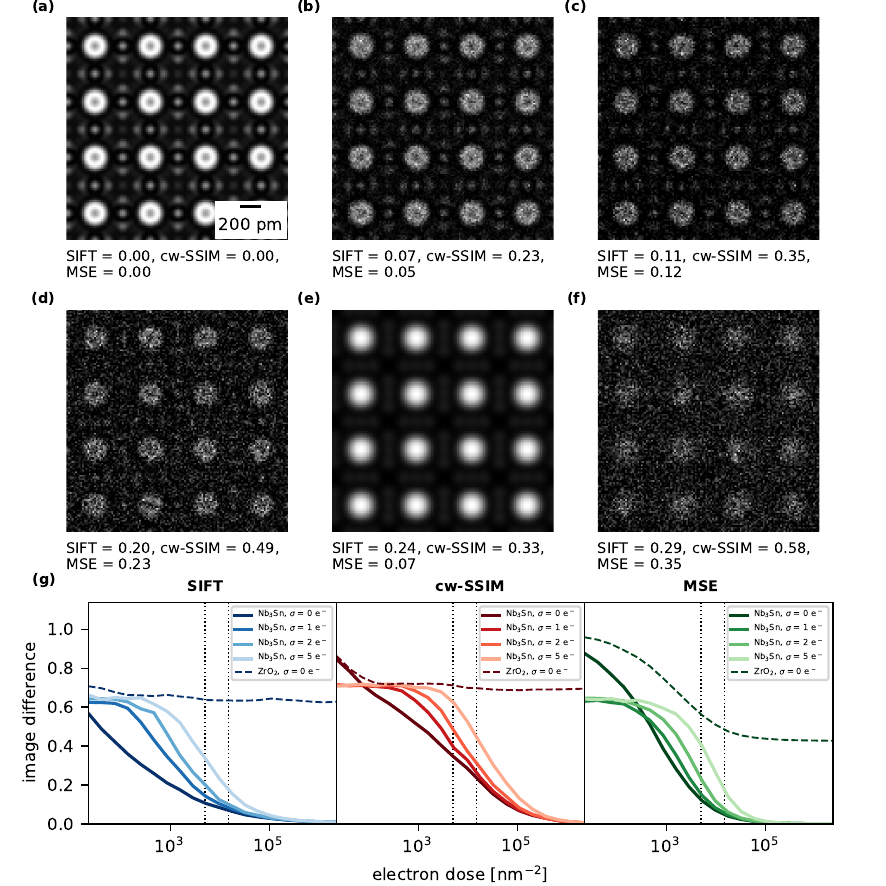}
    \caption{Comparison of image difference measures for various types of noise. 
      (a) Reference image: \ce{Nb3Sn} with a projected sample thickness of 21.1 nm and $128 \times 128$ pixel. 
      (b) Shot noise with 15000 e$^-$/nm$^2$. (c) Shot noise with 5000 e$^-$/nm$^2$. 
      (d) Shot noise with 5000 e$^-$/nm$^2$ with additional Gaussian noise with $\sigma = 2 $ e$^-$.
      (e) Gaussian blurring with $\sigma_{\mathrm{blur}} = 3$ pixel $ = 0.049 $ nm. 
      (f) Noise distortions of (d) and (e) combined.
      (g) Image difference to the reference image as function of the electron dose with additional Gaussian noise.
}
    \label{fig2}
\end{figure*}

We compare the HRTEM image of pristine \ce{Nb3Sn} [Fig.~\ref{fig2}(a)] to various images distorted by noise.
This serves as a way to gauge the realistic conditions necessary in order to still get reliable measures
from the difference metrics.
We have forgone comparison with more classic image distortions like translation, rotation, spatial scaling or a luminance shift,
as one in practice should be able to eliminate these distortions by properly aligning, cutting or normalizing the images.
Instead, we put focus on different kinds of noise pollution.
Fig.~\ref{fig2}(b) and (c) show the $128 \times 128$ pixel reference image with a finite electron dose of 15000 and 5000 e$^-$/nm$^2$, respectively. 
While 15000 e$^-$/nm$^2$ is relatively high compared to the low electron dose regime, the cw-SSIM difference is already higher than the previously mentioned
threshold of discernibility. 
Such high sensitivity to shot noise indicates that cw-SSIM is less suited than the other methods for this particular type of noise, as the perfect image difference metric should be as little as possible influenced by noise and still be able to detect the same underlying crystal structure of both images.
Gaussian noise, used to model read-out noise, electronic noise, etc., results in a strong response for all the metrics [Fig.~\ref{fig2}(d)].
Gaussian blurring, used to model instrumental broadening due to the partial coherence of the electron source \cite{Schattschneider2012}, shows a different trend than the other image modifications.
For Gaussian blurring the MSE image difference is significantly lower compared to the other metrics, making it the first choice when blurring alone is of concern.
When combining shot noise, Gaussian noise and Gaussian blurring [Fig.~\ref{fig2}(f)] it becomes evident that the
image difference contributions of the individual noise types are, at least to some degree and for small image modulations, additive
for all employed metrics.
This is favorable, as Fig.~\ref{fig2}(f) also shows the biggest difference to the reference image using subjective evaluation.\\
We find that for extremely low electron doses (below 100 e$^-$/nm$^2$) the SIFT image difference metric systematically results in a lower difference compared to cw-SSIM and MSE [Fig.~\ref{fig2}(g)].
This behaviour is expected, as the image difference should converge towards the value for the black image [Fig.~\ref{fig_bench}(f)] when the electron dose approaches zero and no other type of noise is present.
When Gaussian noise is added, even an almost negligible amount, all of the applied image difference metrics approach a value around 0.7 with the electron dose approaching zero. 
This specific value corresponds to the image difference of the reference image to an image without structure with only Gaussian noise [Fig.~\ref{fig_bench}(d)].
All three difference metrics assign this case a lower image difference than that of a completely black image, 
hence the curiosity that for low electron doses Gaussian noise results in a lower image difference compared to the 
absence of Gaussian noise.
In order to further gauge the usefulness of the difference metrics in the low electron dose regime we compare the dose dependent difference of \ce{Nb3Sn} to the reference image (same material) to the dose dependent difference of \ce{ZrO2} to the reference image (different material).
When the image difference from noise alone approaches the image difference due to the change of crystal structure the metric can not effectively discern between the materials and is, thus, no longer useful.
According to this, of the three investigated metrics, the metric based on the SIFT algorithm emerges as best suited for extremely low electron doses.

\noindent
The individual performance of the metrics for images heavily modified by noise, especially shot noise, serves not only 
as a way for us to characterize and rank them compared to each other. 
On the contrary, an ever growing interest in the characterization of beam-sensitive materials is emerging \cite{Chen2020}.
These materials include organic materials usually in the context of cryogenic-TEM techniques \cite{Dubochet2018, Fernandez-Leiro2016, Glaeser2018}, 
organic-inorganic hybrid materials such as metal-organic frameworks \cite{Zhu2013, Wiktor2017, Liu2020},
zeolites \cite{Mayoral2011, Ugurlu2011},
and low-dimensional materials \cite{Susi2013, Zan2013, Garcia2014}.
The fragile nature of these materials makes them highly vulnerable to knock-on damage, radiolysis, amorphization or 
decomposition of covalent bonds. 
Thus, only extremely low electron doses (100 - 3000 electrons/nm$^2$) are possible and only images with a low 
signal-to-noise ratio are available.
In order to perform any automated image characterization, detection or image evaluation for these materials,
it is crucial that the used image metric can still produce reliable comparisons under these conditions.

\subsection{Detectability of Precipitates}
\begin{figure*}
    \centering
    \includegraphics{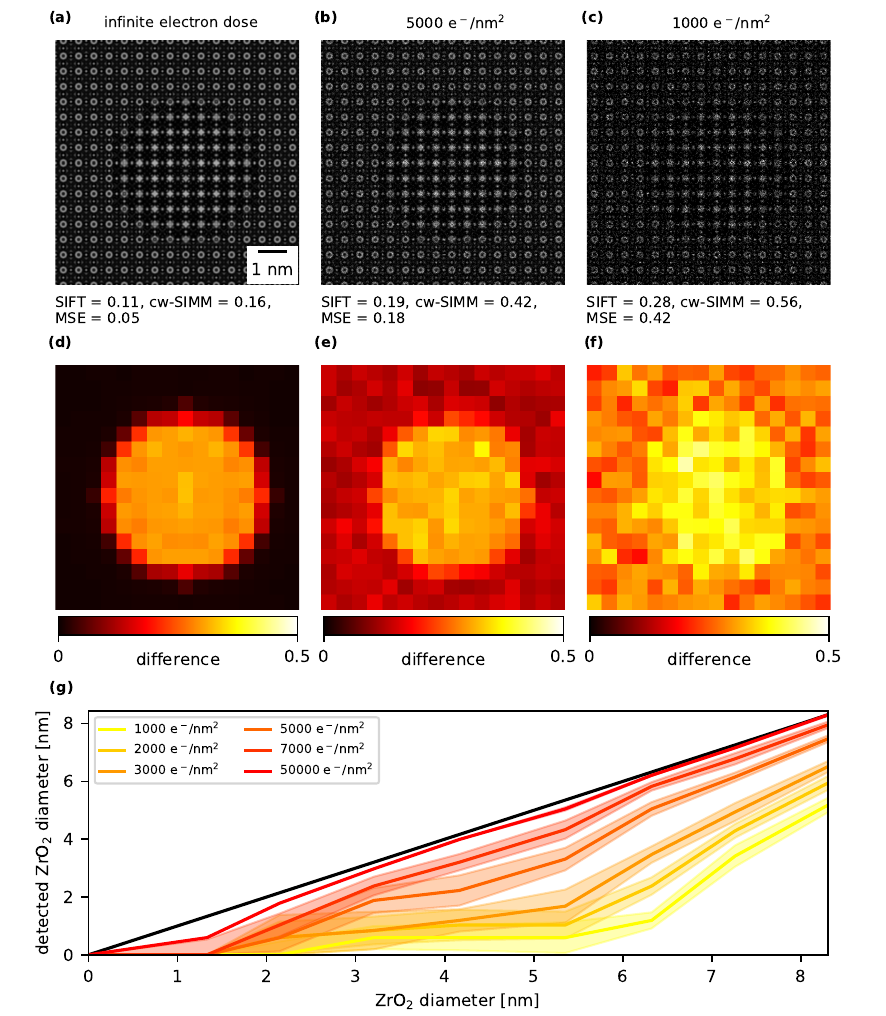}
    \caption{(a), (b), (c) \ce{Nb3Sn} with a spherical precipitate of \ce{ZrO2} with a diameter of 10 unit cells (5.2 nm) in the center, for a total sample thickness of 21.1 nm.
      Image difference calculated relative to the pristine reference image in Fig.~\ref{fig2}(a).
      (d), (e), (f) SIFT image difference of each unit cell of the images in (a), (b) and (c), respectively, to the reference image.
      (g) Detected diameter of the \ce{ZrO2} precipitate determined using the SIFT difference metric for various precipitate radii and electron doses.
      The black line shows the actual precipitate size.
      The filled area around the lines indicates the standard deviation of the detected diameter calculated from repeatedly applying 
      the shot noise.
}
    \label{fig3}
\end{figure*}

\begin{figure}
    \centering
    \includegraphics{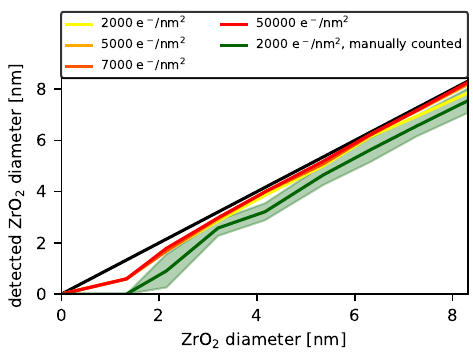}
    \caption{Same initial conditions as in Fig.~\ref{fig3} but with a dose dependent image difference threshold more optimized to the specific problem at hand. In comparison with the detected diameter by manually counting unit cells by subjective evaluation. 
}
    \label{fig_opt}
\end{figure}

In order to demonstrate the usefulness as well as the limits of the chosen difference metrics we use them to determine the size of \ce{ZrO2} precipitates in a \ce{Nb3Sn} substrate.
This model task, while primarily serving as a performance test for the difference metrics, can easily be 
extended to real life applications. 
Thus, this method could automate size determination of various precipitates or nano materials from HRTEM images.
Further and complementary, for a given noise level the size underestimation based on visual inspection can be characterized.\\
We choose \ce{Nb3Sn} as our substrate material for a number of reasons.
\ce{Nb3Sn} is well-known for its superconducting properties. 
It is used in the construction of a new generation of superconducting wires \cite{Ambrosio2015}
with higher critical magnetic fields than previous generations and considered for the high demands of a
post-LHC (Large Hadron Collider) particle accelerator at CERN \cite{Schoerling2019,FCCfuture,Pfeiffer2020}.
As it is a type II superconductor, a large part of its effectiveness is determined by the materials
ability to pin the flux lines, i.e. the normal conducting tubes of magnetic flux through the material
\cite{FluxPinning1974,Matsushita2014,Scanlan1975}.
This, in turn, is presumedly determined by the number and size distribution of nano-precipitates in the material, e.g. of \ce{ZrO2} \cite{Pfeiffer2020}.
Thus, an image analysis method that automatically measures precipitate sizes and over- or underestimation for a given noise level is crucial.\\
In order to demonstrate the feasibility of the approach, we take the pristine \ce{Nb3Sn} crystal from Fig.~\ref{fig2} and replace an increasingly growing sphere of unit cells in the center with \ce{ZrO2} unit cells and subsequently apply shot noise [Fig.~\ref{fig3}(a)-(c)].
The resulting image is then compared individually for each unit cell to the reference image [Fig.~\ref{fig3}(d)-(f)].
We only present results obtained with the SIFT difference metric as it proved to be the most suitable for the task, mainly due to the aforementioned better response to shot noise.
Results for the MSE and cw-SSIM metrics can be found in the SI.
The downside to using the SIFT algorithm is, however, that a large number of pixels in the images is necessary in order to still reliably detect features, as we only compare small fractions of the image.
We choose to simulate the images with $1024 \times 1024$ pixels, resulting in $64 \times 64$ pixels per unit cell.
This number appears to be sufficiently high while still keeping computation times manageable. 
Note, however, when shot noise is involved the appearance of an image can vary strongly with the number of pixels despite having
the same electron dose.\\
The unit cell resolved difference map for infinite electron dose [Fig.~\ref{fig3}(d)] shows a high difference in the center of the \ce{ZrO2} circle and a lower difference at the edge due to the decreasing projected thickness of the precipitate.
Nevertheless, there is a clear cut between \ce{Nb3Sn} and \ce{ZrO2}, more so than the HRTEM image would indicate.
An image difference threshold can be found so that the algorithm detects each replaced unit cell and thus the exact precipitate size.
For an infinite electron dose and the SIFT algorithm we have chosen a difference of 0.18 as the threshold.
Upon decreasing the electron dose the shot noise results in a background difference for the non-replaced \ce{Nb3Sn}
unit cells compared to the reference image with infinite dose.
The average difference due to shot noise alone is determined and added to the threshold for automatic unit cell counting,
allowing us to still detect the \ce{ZrO2} unit cells in the center of the precipitate without wrongly counting unreplaced, albeit noisy, \ce{Nb3Sn} unit cells.
The clear cut between base material and precipitate, however, vanishes for increasing shot noise.
Generally, this leads to an underestimation of the counted unit cells and subsequently the determined
precipitate size.
Further, when the electron dose gets too low or the precipitate too small, the image difference from the \ce{ZrO2} unit cells can no longer be distinguished from the image difference background due to noise.
In this case the detection algorithm is no longer valid and no reliable particle diameter can be determined.
Fig.~\ref{fig3}(g) indicates that above an electron dose dependent particle size the detected diameter increases linearly with increasing diameter. 
Thus, the detected precipitate diameter is underestimated by an approximately constant value for a given electron dose.\\
\noindent
The error in counting for lower doses is not primarily a result of the inability of the SIFT image difference metrics to distinguish between the different crystal structures, as it would be for the cw-SSIM metric (see Fig.~S3). 
It is, however, a consequence of the chosen threshold and, in turn, the algorithm responsible for counting the unit cells which we wanted to keep as general as possible.
For completeness, we present the detected precipitate diameters when an electron dose dependent threshold function more optimized to the specific problem is used (Fig.~\ref{fig_opt}).
The new threshold function is an empirically found combination of the base threshold, the average difference due to shot noise and the contrast in the image difference map. 
While the detected \ce{ZrO2} diameters also generally underestimate the true value, the error is negligible except for very small diameters or electron doses. 
Further improvement could still be achieved by implementing \textit{a priori} knowledge of the precipitate shape or by not counting single \ce{ZrO2} unit cells.
The results, however, already surpass the method of manually counting unit cells by accuracy and, more important, by efficiency and time duration of the procedure.

\noindent
While all three of the chosen image difference methods can be successfully used for automatic detection or quantification of certain features in HRTEM images, SIFT has yielded the best results for this specific task involving low electron doses.
As the SIFT method only takes key features into account and not the visual structure of the whole image it is more resilient to shot noise compared to the other two methods.
None of the three methods, however, has emerged as the best suited for all type of images, as they all have situations in which one works better or worse.
When images modified by various types of noise are used and compared to the pristine reference image, all three methods show characteristics comparable to those of subjective human image evaluation, which is one of the main requirements of an image difference metric.
The effectiveness of a method here is not determined by how well it can evaluate the image quality but by how well 
it can ignore the noise and compare the underlying images.

\section{\label{Conclusion}Conclusion}
In this work we have examined the feasibility of applying difference metrics to tasks involving HRTEM images.
We find that the three chosen methods, SIFT, cw-SSIM and MSE, while in general providing similar results,
all have different areas where one outshines the others.
With this in mind we have successfully applied the SIFT metric to the task of automatically detecting the size of \ce{ZrO2} precipitation particles in \ce{Nb3Sn}.
In the future we plan on using the investigated image difference metrics on energy-filtered maps
with the intent of parameter optimisation for orbital mapping \cite{Loeffler2017},
where an automatic image analysis method can replace the need of manually investigating thousands of images.\\

\noindent
We acknowledge financial support by the Austrian Science Fund (FWF) under grant number I4309-N36.
Further, we acknowledge TU Wien Bibliothek for financial support through its Open Access Funding Programme.\\

\bibliography{metric, TEM}
\bibliographystyle{elsarticle-num}

\end{document}


\begin{frontmatter}

\title{Supplementary Information for: "Image Difference Metrics for High-Resolution Electron Microscopy"}

\author[1]{Manuel Ederer}
\ead{manuel.ederer@tuwien.ac.at}

\author[1]{Stefan L\"{o}ffler}
\address[1]{University  Service  Centre  for  Transmission  Electron  Microscopy, TU Wien, Wiedner Hauptstraße 8-10/E057-02, 1040 Wien, Austria}

\date{\today}

\end{frontmatter}

\section*{Image difference algorithms}
\subsection*{SSIM}
\noindent
In this section we present the structural similarity index measure from \cite{wang2004} and \cite{wang2005} in more detail including all parameters used in our calculations.
For each pair of points $\mathbf{x}$ from image A and $\mathbf{y}$ from image B a local batch of pixels of equal size and shape around the points is taken.
On this batch the luminance is compared
\begin{equation}
    l(\mathbf{x}, \mathbf{y}) = \frac{2\mu_x\mu_y + C_1}{\mu^2_x + \mu_y^2 + C_1},
\label{eq:luminance}
\end{equation}
\noindent 
consisting of the local image signal mean
\begin{equation}
    \mu_x = \frac{1}{N} \sum_{i=1}^{N} \mathbf{x}_i,
\end{equation}
where $\mathbf{x}_i$ are the points of the batch around $\mathbf{x}$ and $C_1$ is a small constant to avoid instability when $\mu_x^2 + \mu_y^2$ is close to zero.
In the next step the local mean is removed from the signal and subsequently the contrast is compared.
The contrast comparison function 
\begin{equation}
    c(\mathbf{x}, \mathbf{y}) = \frac{2\sigma_x\sigma_y + C_2}{\sigma_x + \sigma_y + C_2}
\label{eq:contrast}
\end{equation}
takes a similar form to the luminance comparison function Eq.~\ref{eq:luminance} with the local signal variance 
\begin{equation}
    \sigma_x = \frac{1}{N-1} \sum_{i=1}^N (\mathbf{x}_i - \mu_x)^2 
\end{equation}
\noindent
and a small constant $C_2$.
In the next step of the workflow the image signal is normalised by its own standard deviation.
Lastly, the structure comparison function is defined as
\begin{equation}
    s(\mathbf{x}, \mathbf{y}) = \frac{\sigma_{xy} + C_3}{\sigma_x\sigma_y + C_3}
\label{eq:structure}
\end{equation}
\noindent
with the local correlation coefficient $\sigma_{xy}$ between $\mathbf{x}$ and $\mathbf{y}$ 
\begin{equation}
    \sigma_{xy} = \frac{1}{N-1} \sum_{i=1}^N (\mathbf{x}_i-\mu_x) (\mathbf{y}_i-\mu_y)
\end{equation}
and a small constant $C_3$.
Combining all three comparison functions of Eq.~\ref{eq:luminance}, Eq.~\ref{eq:contrast} and Eq.~\ref{eq:structure} and choosing $C_3 = C_2/2$ results in the original form of the SSIM index

\begin{equation}
    \begin{split}
        SSIM&(\mathbf{x}, \mathbf{y}) = l(\mathbf{x}, \mathbf{y}) c(\mathbf{x}, \mathbf{y}) s(\mathbf{x}, \mathbf{y}) = \\
        & = \frac{(2\mu_x \mu_y + C_1)(2\sigma_{xy} + C_2)}{(\mu_x^2 + \mu_y^2 + C_1)(\sigma_x + \sigma_y + C_2)}.
    \end{split}
    \label{eq:SSIM_app}
\end{equation}

\noindent
In order to achieve more robustness against image noise, the similarity index is extended to the complex wavelet transform domain \cite{wang2005}.
The continuous complex wavelet transform of a real signal $f(\mathbf{x})$ is given by

\begin{equation}
    F(s,\mathbf{p}) = \frac{1}{|s|^{1/2}} \int_{-\infty}^{\infty} f(\mathbf{x}) \psi \left( \frac{\mathbf{x}-\mathbf{p}}{s} \right)^{\ast} \del \mathbf{x} 
\end{equation}
where $\psi(\mathbf{x})$ is a continuous, complex function called the mother wavelet from which the daughter wavelets are constructed by shifting and scaling. 
In our case the signal $f(\mathbf{x})$ represents the image and the 2-dimensional parameter $\mathbf{x}$ the pixel positions. 
The position parameter $\mathbf{p}$ is used to scan over the signal in pixel space and in the discrete case has the same range as the pixels of the image. 
Through the scale parameter $s$ the wavelet transform scans the signal over multiple (spatial) frequencies. 
We choose s from the set $\{s | s = 1, ..., 30; ~s \in \mathbb{N}\}$.
We use the complex Morlet wavelet \cite{Ashmead2012} (or Gabor wavelet) defined by
\begin{equation}
    \psi(x) = \pi^{-\frac{1}{4}} \mathrm{e}^{-\frac{1}{2} x^2} \mathrm{e}^{\mathrm{i}\omega_{0}x}
\end{equation}
where $\omega_0$ is the center frequency of the wavelet.
For each corresponding point $\mathbf{p}_{0}$ of the wavelet transformed images being compared, the 30 coefficients $c_i = F(s, \mathbf{p}_{0})$ are used for Eq.~\ref{eq:SSIM_app}.
Instead of a local batch of pixels in real space, a set of coefficients of the wavelet transform is used for the calculation of mean, standard deviation and correlation coefficient.
Due to the bandpass nature of the wavelet filters the coefficients have zero mean, resulting in the original form of the complex wavelet SSIM
\begin{equation}
    \Tilde{D}(\mathbf{x}, \mathbf{y}) = \frac{2 | \sum_{i=1}^N c_{\mathbf{x},i} c^*_{\mathbf{y},i}| + K}{\sum_{i=1}^N | c_{\mathbf{x},i}| ^2 + \sum_{i=1}^N |c_{\mathbf{y},i}|^2 + K}
\end{equation}
where $N=30$ is the number of coefficients and K is a small, positive constant.
Subsequent averaging over all points leads to a total image similarity measure, exactly as in the real space case.

\subsection*{SIFT}
\noindent
The image difference based on the scale-invariant feature transform (SIFT) \cite{lowe1999} is calculated according to the workflow diagram in Fig.~\ref{SI_fig2}.

\renewcommand{\thefigure}{S1}
\begin{figure*}
    \centering
    \includegraphics[width=16.5cm]{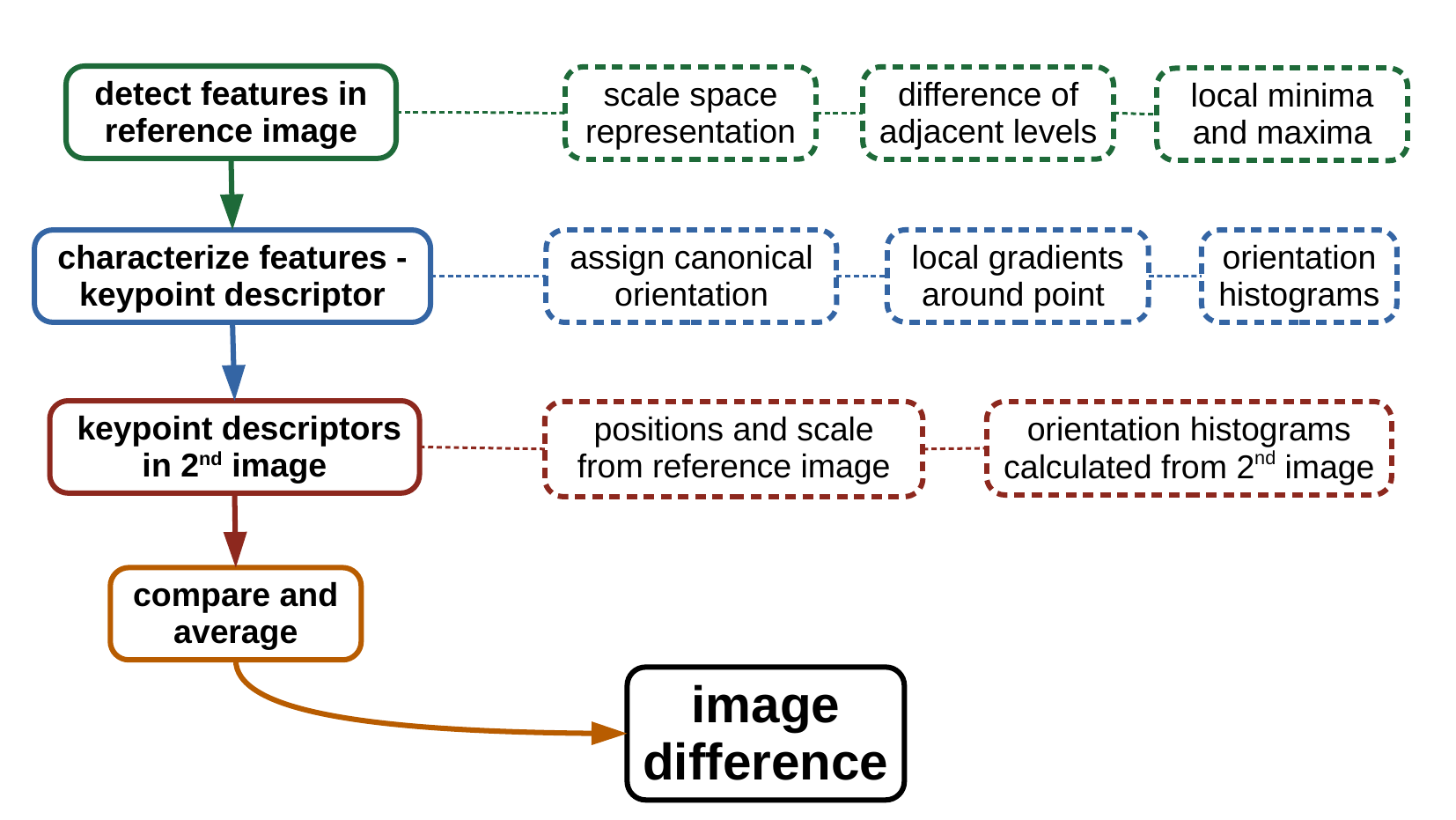}
    \caption{Workflow diagram of the SIFT measurement
}
    \label{SI_fig2}
\end{figure*}
\noindent
The first step consists of finding features in the reference image that are independent of global translations and rotations, noise and scaling of the image.
In order to ensure the last point an image pyramid in scale space \cite{Witkins1983} is constructed. 
Each level, or scale, of the pyramid is an increasingly more blurred version of the original image.
For a certain scale $\sigma$ the scale-space representation $L(x,y,\sigma)$ of a 2-dimensional signal $I(x,y)$ is given by
\begin{equation}
    L(x,y,\sigma) = G(x,y,\sigma) * I(x,y),
\end{equation}
where $*$ is the convolution operation in $x$ and $y$, and 
\begin{equation}
    G(x,y,\sigma) = \frac{1}{2\pi\sigma^2}e^{-(x^2 + y^2)/2\sigma^2}
\end{equation}
is the Gaussian function.
After an octave, i.e. the doubling of $\sigma$, the Gaussian image is resampled by a factor of 2.
In each octave the difference between images of scales separated by a constant multiplicative factor k is calculated, 
resulting in the difference-of-Gaussian (DoG) function
\begin{equation}
    D(x,y,\sigma) = L(x,y,k\sigma) - L(x,y,\sigma).
\end{equation}
Blurring an image with a Gaussian kernel suppresses only high-frequency spatial information. 
Thus, the DoG acts like a band-pass filter, attenuating spatial frequencies outside of the range between $\sigma$ and $k\sigma$.
The pyramid of Gaussians and DoG can be schematically seen in Fig.~\ref{SI_fig3}.
\renewcommand{\thefigure}{S2}
\begin{figure*}
    \centering
    \includegraphics[width=15cm]{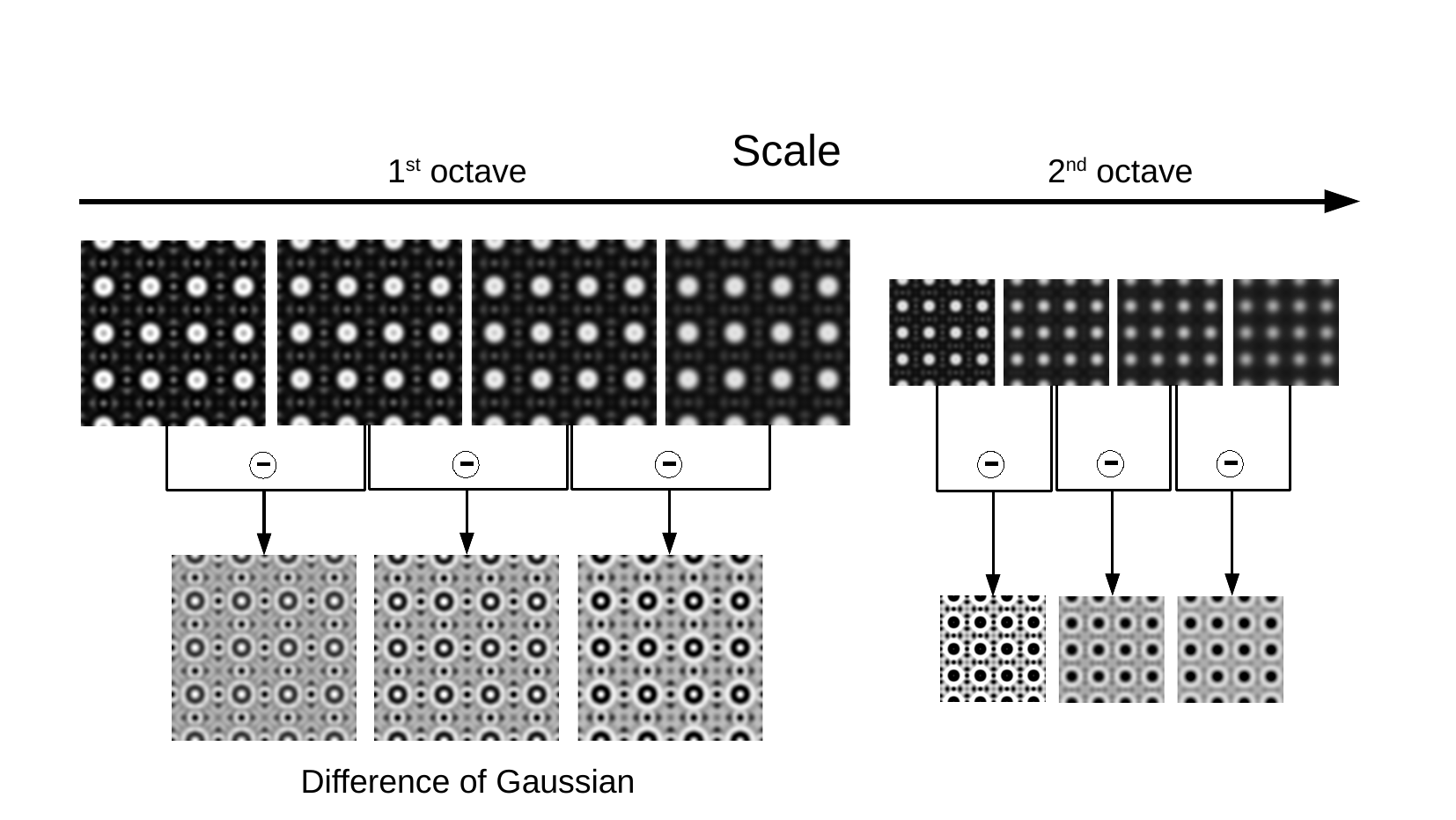}
    \caption{Pyramid of Gaussians in scale-space.
    Top row: each image is the result of consecutively more blurring of the original image.
    Neighbouring images are separated by a constant factor in scale-space.
    After an octave (doubling of $\sigma$) the image is resampled by a factor of two. 
    In each octave neighbouring scales are subtracted from each other resulting in the pyramid of difference-of-Gaussian functions in the bottom row.
    Each pixel is then compared to its 8 neighbours at the current scale and its 9 neighbours at the neighbouring scales each in order to find local maxima and minima.
}
    \label{SI_fig3}
\end{figure*}
\noindent
In this work we have used 3 layers of DoG per octave which is also the value used in \cite{Lowe2004}.
Local maxima and minima of the DoG images are detected by comparing each pixel to its 8 neighbours at the same scale and to its 
9 neighbours each in the scale above and below.
Afterwards, points located in areas of low contrast or along edges are discarded.
The remaining keypoints are locations in scale-space marking features that are distinctive for the image.\\
The next step of the SIFT image difference algorithm consists of characterizing the found features, i.e. in constructing the keypoint descriptors.
In order to assign a canonical orientation to the keypoint found at scale $\sigma_0$, a small circular batch of pixels around the point in the Gaussian smoothed image $L(x,y,\sigma_0)$ is taken into account.
For each pixel of this batch the local gradient magnitude $m(x,y)$ and gradient orientation $\theta(x,y)$ is calculated using pixel differences:
\begin{equation}
    \begin{split}
        m(x,y)^2 = & (L(x+1,y)-L(x-1,y))^2 ~ + \\ 
        & + (L(x,y+1)-L(x,y-1))^2
    \end{split}
\end{equation}
\begin{equation}
    \theta(x,y) = \arctan \left( \frac{L(x,y+1)-L(x,y-1)}{L(x+1,y)-L(x-1,y)} \right).
\end{equation}
\noindent
The local gradient angles $\theta(x,y)$ are distributed to a histogram of 36 angle bins and weighted with the respective magnitude $m(x,y)$ and by a Gaussian-weighted circular window.
The highest peak of this histogram is then taken as the canonical orientation of the keypoint and all subsequent description will be relative to this orientation.
For the descriptor itself a 16 $\times$ 16 pixel window around the 
keypoint of the smoothed image $L(x,y,\sigma_0)$ is used.
The area is partitioned into 16 squares of 4 $\times$ 4 pixels and the remaining gradients are calculated, if they have not been already calculated in the previous step.
In each of the 16 partitions the gradient orientation values are inserted into histograms with 8 orientation bins.
In order to avoid too abrupt descriptor changes trilinear interpolation is used to distribute the values among an orientation bin and its adjacent bins.
All the histograms from the partitions together then form the keypoint descriptor vector with 16 $\times$ 8 = 128 elements.
The third step of the SIFT image difference workflow differs significantly from the original algorithm outlined in \cite{lowe1999, Lowe2004}.
In the second image we do not again detect features as described in the first step, instead we take the same positions and scale of the keypoints in the reference image.
The areas around the positions are, however, taken from the (smoothed) second image and the resulting orientation histograms and descriptor vectors as well. 
Thus, for each feature in the reference image we arrive at a descriptor vector from the reference image and one from the second image, which are to be compared.
In the final step, we take the Frobenius norm of the difference between each pair of descriptor vectors and average them, resulting in a total image difference.


\section*{Precipitate size determination}
\noindent
In this section we present the results of the \ce{ZrO2} precipitate size determination using the cw-SSIM and the MSE image difference metric.

\renewcommand{\thefigure}{S3}
\begin{figure*}
    \centering
    \includegraphics{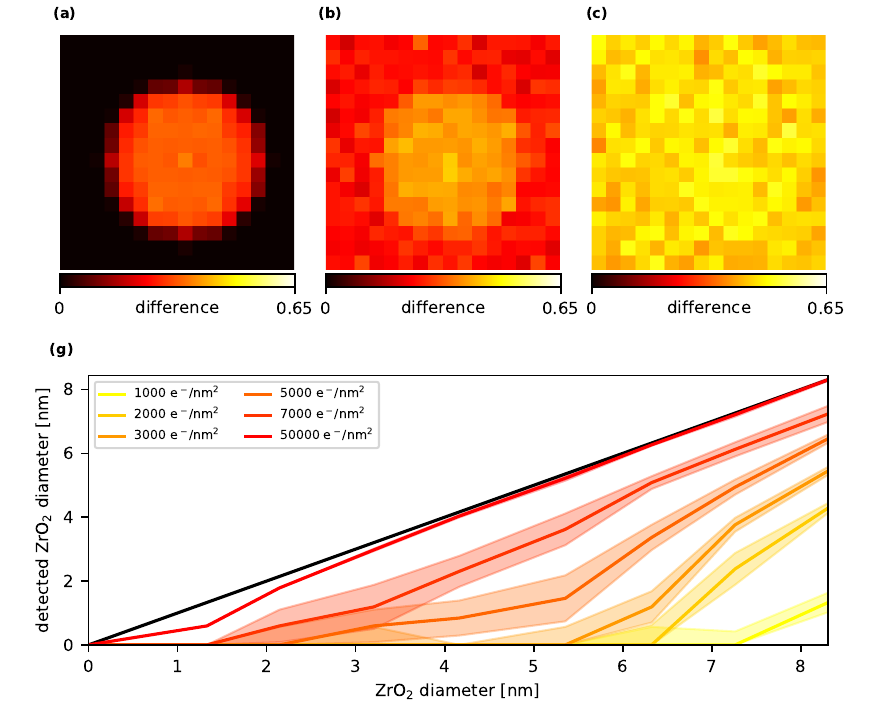}
    \caption{Same as Fig.~4 but with the cw-SSIM metric.
}
    \label{SI_SSIM}
\end{figure*}

\renewcommand{\thefigure}{S4}
\begin{figure*}
    \centering
    \includegraphics{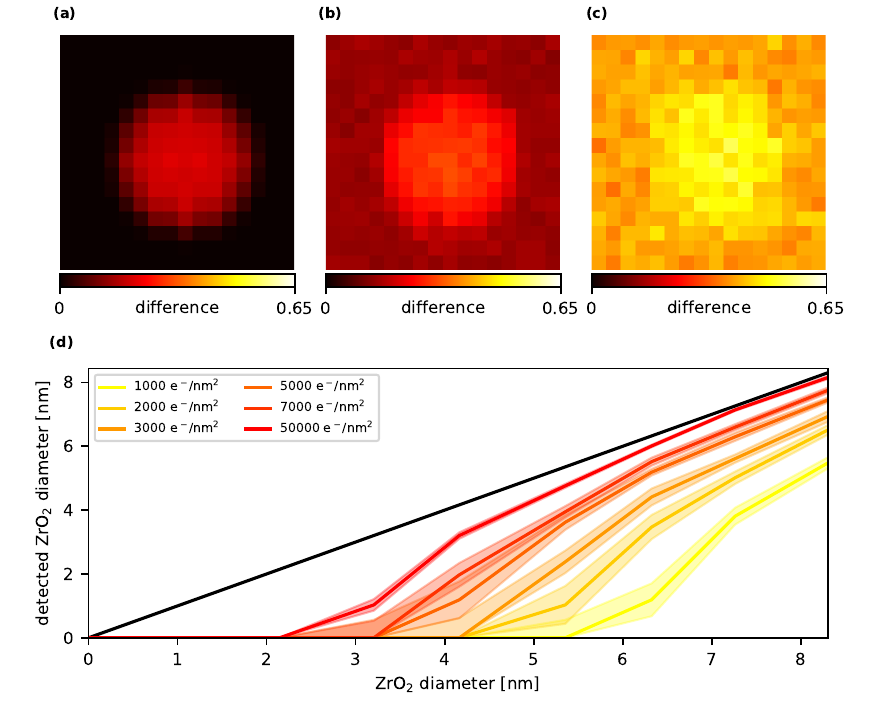}
    \caption{Same as Fig.~4 but with the MSE metric.
}
    \label{SI_MSE}
\end{figure*}

\bibliography{metric, TEM}
\bibliographystyle{elsarticle-num}